\documentclass
[aps,prl,twocolumn,floats,balancelastpage,showpacs,amssymb,floatfix,nofootinbib,superscriptaddress,amsmath]{revtex4-1}

\usepackage{epsfig}
\usepackage{epstopdf}

\usepackage{float}
\usepackage[usenames]{color}
\usepackage{color,epsfig,amsmath,amssymb,fancyhdr}   
\usepackage[french]{babel}
\usepackage[T1]{fontenc}
\usepackage{ulem}

\begin{document}

\title{
Direct Searches for Hidden-Photon Dark Matter with the SHUKET Experiment
}

\author{Pierre Brun}
\author{Laurent Chevalier}
\author{Christophe Flouzat}

\affiliation{Irfu, CEA, Universit\'e Paris-Saclay, F-91191 Gif-sur-Yvette, France}

\begin{abstract}
Hidden photons are dark matter candidates motivated by theories beyond the standard model of particle physics. They mix with conventional photons, and they can be  detected through the very weak electromagnetic radiation they induce at the interface between a metal and the air. SHUKET (SearcH for U(1) darK matter with an Electromagnetic Telescope) is a dedicated experiment sensitive to the hidden photon-induced signal. The results from a hidden photon search campaign are reported, with no significant detection of a dark matter signal. Exclusion limits are derived from the observed noise fluctuations in a 5~GHz to 6.8~GHz frequency range, corresponding to a hidden photon mass region ranging from 20.8~$\mu$eV to  28.3~$\mu$eV. SHUKET is currently the most sensitive instrument in this mass range and the obtained limits on the kinetic mixing term constrain significantly dark matter models inspired from string theory.
\end{abstract}


\maketitle

{\it Introduction --}A fraction of the dark matter (DM), if not all of it, could have been produced in the early universe through non-thermal processes. A common example of such a process is the misalignment mechanism, which can be summarized as follows. At the end of inflation, fields are expected to have a random initial value. When the Hubble rate becomes comparable to the mass of the field, the latter starts oscillating to minimize its potential. If its decay rate is low, the oscillations behave as cold DM. Indeed the energy density scales as the inverse third power of the scale factor and, as a consequence of equipartition, the field is pressureless. The misalignment mechanism is relevant in particular in the much-studied case of axion DM. There, the initial condition is picked out during the breakdown of the Peccei-Quinn symmetry and the axion field starts oscillating after it acquires its mass during the quark-hadron phase transition. The same mechanism applies for hidden photons (HP).

In this letter, we report the results from a search for HP. They are hypothetical new gauge bosons associated to a new U(1) symmetry for which standard particles carry no charge. Such additional Abelian symmetries are quite common in string theory. The dominant interaction between the visible and the hidden sector is realized through the kinetic mixing between photons $A^\mu$ and HP~\cite{2011PhRvD..84j3501N,  2004PhRvD..70k5009M}. In the absence of sources, the relevant part of the Lagrangian is 
\begin{equation}
\mathcal{L} = -\frac{1}{4}F_{\mu\nu}F^{\mu\nu}-\frac{1}{4}\phi_{\mu\nu}\phi^{\mu\nu}-\frac{m^2}{2}\phi_\mu\phi^\mu - \frac{\chi}{2}F_{\mu\nu}\phi^{\mu\nu}
\end{equation}
where $F^{\mu\nu}$ is the field strength of the ordinary photon, $\phi^\mu$ is the massive HP of mass $m$ and $\phi^{\mu\nu}$ its field strength. The dimensionless parameter $\chi$ is the mixing parameter, the value of which can be predicted in the framework of some string models. Depending on the implementation of the models, its value could range from $10^{-12}$ to $10^{-3}$~\cite{1997NuPhB.492..104D,2009JHEP...11..027G, 2012JHEP...01..021G}. In this letter, we assume that part of the DM is in the form of a HP field, produced non-thermally {\it via} the misalignment mechanism~\cite{2011PhRvD..84j3501N,  2004PhRvD..70k5009M}. As a consequence, the corresponding local DM density is made of the energy density of the oscillations of the HP field. At first order, one can describe the photon-HP system as a spatially constant mode of pulsation $\omega = m $. Due to the mixing with the conventional photon, the local DM density induces a small electric field $\vec{E}_{\rm DM} = \chi\, \omega \, \vec{\phi}_{\rm DM}$, which oscillates at the same pulsation $\omega$. The first order correction to this value is due to the velocity of the Earth in the DM halo and the velocity dispersion of the DM gas and is $v^2/2$. Using a typical value $v=10^{-3}$, one can see that the dispersion around the central frequency is of the order of $10^{-6}$. In this work, we analyze the results of the SHUKET experiment (SearcH for U(1) darK matter with an Electromagnetic Telescope), that aims at detecting this weak DM-induced electric field.

\begin{figure}[h]
\includegraphics[width=.8\columnwidth]{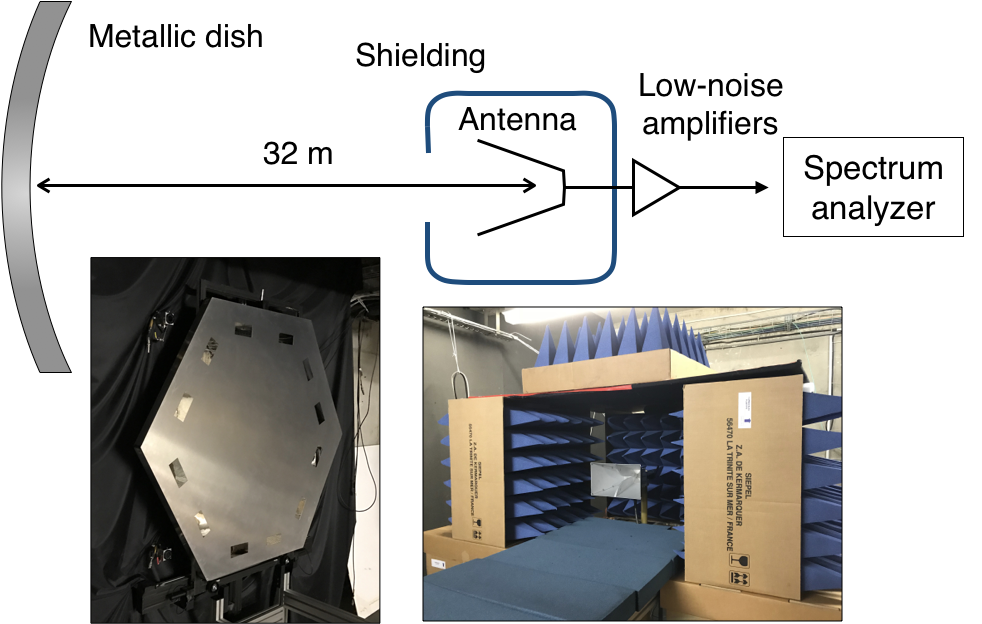}
\caption{Top : Principle of the SHUKET experiment. Bottom left: Picture of the spherical cap, reflective patches are to help with the alignment procedure. Bottom right: Horn antenna with the shielding.\label{fig:1}}
\end{figure}

{\it The experiment --}
The principle of the SHUKET experiment is to apply a boundary condition to the electric field on a surface such that its oscillation generates an outgoing wave. This condition is enforced with a metallic dish, at the surface of which $\vec{E}_{\rm DM, \,\parallel}=\vec{0}$, inducing the emission of an electromagnetic wave perpendicular to the surface. The detection setup is sensitive in the gigahertz range, and the metal used is 3~mm-thick aluminum, so the skin depth is negligible and the above condition applies. The shape of the dish is spherical, so all the emitted waves converge towards a single point. This approach has been proposed in~\cite{2013JCAP...04..016H} as a mean to search for axions and HP. In the case of the SHUKET setup, the dish is a spherical cap of $R=$32~m radius, with an area of 1.2 $\rm m^2$. The principle of the experiment is sketched in Fig.~\ref{fig:1}, together with pictures of the different parts.

Let us now estimate the expected power emitted from the spherical cap. In this analysis, the HP are considered as gas of particles with random directions. Even if the corresponding field has been produced initially in the same state on a spatial scale larger than the apparatus, it is likely that structure formation have randomized their directions afterwards. So the emission is assumed to be unpolarized. The results presented below include upper limits on the power within some frequency bands, so the reader can relax this hypothesis and easily translate our results using other assumptions. The emitted power per unit surface is given by $\langle \vec{E}_{\rm DM}^2 \rangle$ evaluated at the metal/air interface. With $\rho = m^2\langle |\vec{\phi}_{\rm DM} |^2 \rangle/2$ being the local DM density, one finds the total emitted power from the spherical cap :
\begin{equation}
\frac{P}{1\; \rm W}=2.31\times 10^{-20} \left ( \frac{\chi}{10^{-12}} \right )^2 \left ( \frac{\rho}{0.3\;\rm GeV/cm^3} \right )\;\;. \label{eq:power}
\end{equation}
This formula includes a factor 2/3 due to the average over the random orientations of the HP. The power expressed in Eq.~\ref{eq:power} is emitted at a central frequency
\begin{equation}
f =0.24\;{\rm GHz} \times (m/ {\rm \mu eV})\;\;,
\end{equation}
with a modulation of $\delta f/f =5\times10^{-7}$ that is assumed gaussian. The signal is expected to be constant over time. At the center of the sphere, a calibrated horn antenna is placed and aligned. The alignment is performed using optical light emitted slightly off the center of the sphere and observed in the same plane, following the method described in~\cite{2013NIMPA.714...58B}. The same procedure is used to determine the position of the center of the sphere, where the antenna is placed.

The spherical cap is built on the same principle as mirrors for gamma-ray astronomy described in~\cite{2013NIMPA.714...58B}, where the glass has been replaced with aluminum. The sphericity is estimated using visible light; we measured a spread at the center of the sphere of the order of 1 mrad, meaning that all the emission from the dish is contained at its center in a disk of $\sim$3 cm radius. For centimetric waves, the requirement on the sphericity is of course less stringent and this shows the dish can be considered a perfect sphere in our experiment. At 6~GHz the effective area of the antenna is of the order of 50 $\rm cm^2$, so all the emission from the spherical cap is contained. As the expected signal from the spherical dish is a convergent spherical wave, no difference in optical path is expected over the whole surface so diffraction effects are negligible. As shown in~\cite{2016JCAP...01..005J} and~\cite{2013JCAP...11..016J}, the small velocity of the experiment with respect to the DM halo implies an angular shift of the signal. This is however only a few microns in the case of SHUKET so again, the whole signal is concentrated in our antenna. The latter  measures one polarization and we make the conservative assumption that there is no power from the other polarization. The signal being unpolarized under our assumptions, the antenna will collect half the power emitted from the dish.

 The whole setup has been built in a basement on the CEA Paris-Saclay campus, where electromagnetic nuisance is relatively low at the studied frequencies. The antenna is shielded with metallic foam, and a background subtraction is performed as described below. The antenna is a calibrated double ridged horn sensitive between 1~GHz and 18~GHz. The lobe of the antenna is characterized by the manufacturer and in the band relevant to this analysis, all the electromagnetic signal in one polarization from the spherical cap is converted to output power. The field of view of the antenna include regions around the spherical cap, so part of the measured power come from the thermal emission of concrete walls. Estimates of the thermal emission of the mirror itself  show it is negligible at the level of sensitivity of SHUKET. To minimize the influence of parasitic power, the measurements are performed in two steps, with the antenna at the center of the metallic dish ($ON$ runs) and with the antenna a few meters off-axis ($OFF$ runs). In practice the antenna is left in place and the spherical cap is tilted to displace its center. The power measurements from $ON$ and $OFF$ runs are subtracted in the data analysis phase. Note that distant sources of power have their signal focused in the focal plane of the dish, at half its radius, thus far from our antenna. In addition, it has been checked that a signal at the relevant frequencies is effectively detected by our receiver when power is emitted from the focal point at $R/2$ towards the dish. The signal is not detected when the antenna is off-axis in the $OFF$ position.
 
The antenna is connected to a Rohde \& Schwarz FSW signal and spectrum analyzer through a low-noise amplifier (LNA) at ambient temperature. The wide-band state-of-the-art LNA (2.4~dB Noise Figure - DC-7 GHz Bandwidth) amplifies the signal out of the antenna by a factor of the order of 29~dB, such that both signal and noise received by the antenna are 10~dB above the noise floor of the spectrum analyzer.  This way the signal to noise ratio (SNR) of the experiment is preserved before successive or parallel power spectral analysis on 50 MHz bands with a 1~Hz resolution. The precise value of the amplification is  determined with gain calibration runs, it spans between 27.5~dB and 30.5~dB. The LNA being limited to 7~GHz, we explored the highest possible frequencies. Due to the time limitation of the spectrum analyzer loan, we scanned frequencies down to 5~GHz only.

\begin{figure}[h]
\includegraphics[width=.8\columnwidth]{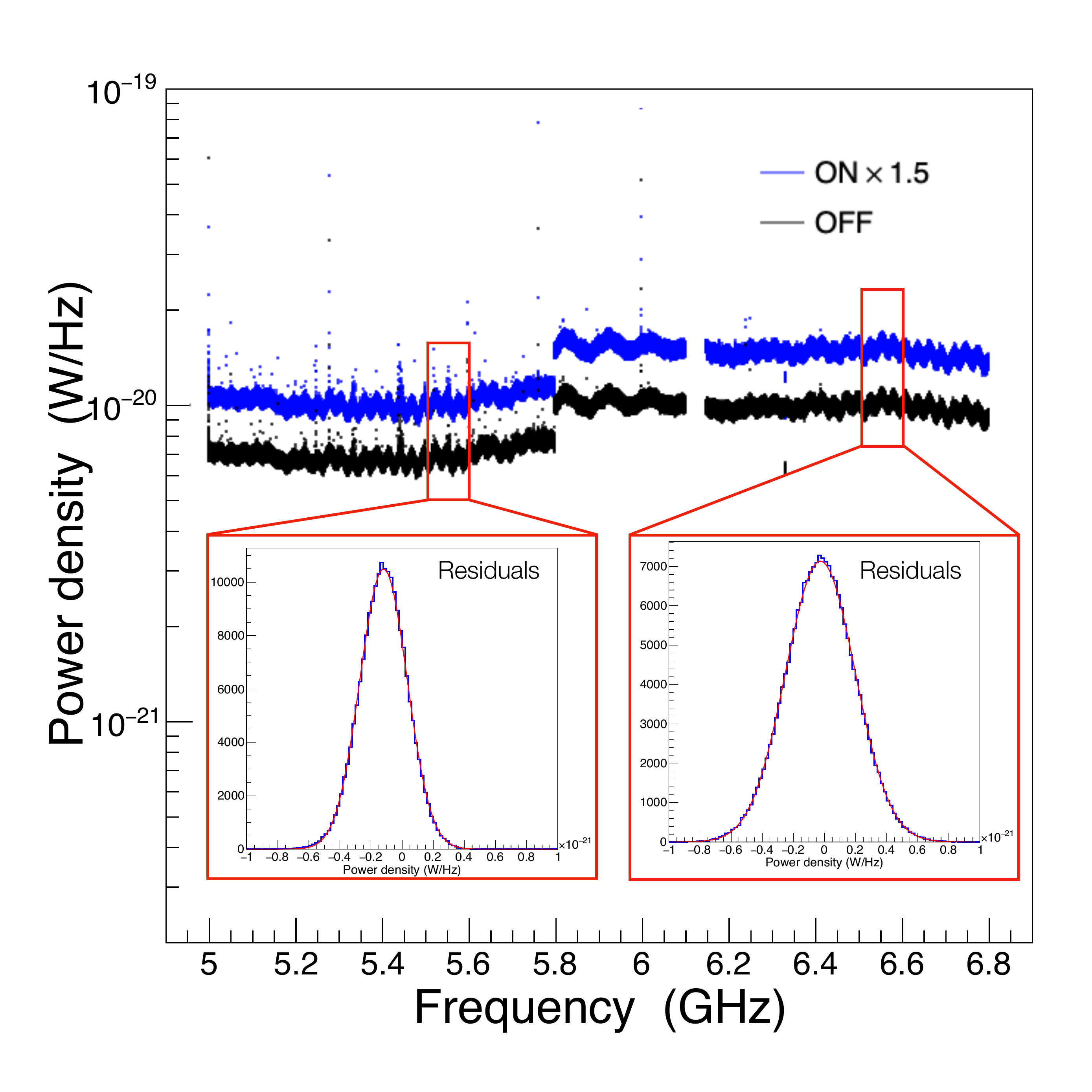}
\caption{Power density measured in the SHUKET experiment before cleaning. The $ON$ distribution is shifted upwards by a factor 1.5 to improve readability. The insets show two distributions of residuals ($ON$-$OFF$) on 100 MHz bands between 5.5~GHz and 5.6~GHz (left) and 6.5~GHz and 6.6~GHz (right).\label{fig:2}}
\end{figure}

{\it Data analysis --}
The data used in this analysis correspond to 120 hours divided in two sets of 97.8 hours and 22.2 hours of $ON$ runs. Each run spans over 50 MHz and last 8000 seconds. The same amount of data has been taken in $OFF$ mode. This dataset corresponds to a scan in frequency between 5~GHz and 6.8~GHz. In addition, a 22.2 hour run  on a single 50 kHz band has been performed to demonstrate an increase in sensitivity is feasible. The cleaning of the data consists essentially in identifying frequency bins in which an excess power is observed in the $ON$ power distribution with respect to the typical fluctuation. Once such a frequency is found, a similar signal is searched for in the $OFF$ distribution. All $ON$ excesses had counterparts in the $OFF$, and no unaccounted significant excess was found. When a frequency bin showed an excess power in both datasets, it is naturally removed when $ON$ and $OFF$ distributions are subtracted. The measured distributions of power density are displayed in Fig.~\ref{fig:2}, where the $ON$ distribution is shifted upwards by a factor 1.5 to improve readability, otherwise the $ON$ and $OFF$ graphs are superimposed. The distributions show the above-mentioned bins with excess power. A modulation is present with respect to the frequency, this is due to the value of the gains of the amplifiers that depend on the frequency. The discontinuity at 5.8~GHz corresponds to a transition between two data taking periods with a few days without data taking during which the power supply for the amplifiers is likely to have drifted. The hole in the data between 6.1~GHz and 6.15~GHz is due to parasitic signals, the corresponding data is dismissed. 

 After cleaning and background subtraction, the overall (5~GHz to 6.8~GHz) residual distribution has a RMS of $2\times 10^{-22}\;\rm W/Hz$. It is not expected to be gaussian on such a large band due to gain fluctuations. However, on narrower scales, the distribution of residuals is gaussian. As an illustration, the insets in Fig.~\ref{fig:2} display the residual distributions in two typical bands of 100 MHz width, together with a gaussian fit. The dispersion of the residuals is compatible with what is expected from the Dicke radiometer formula~\cite{Dicke}: assuming no gain fluctuation,  the detectable power is $P=k_{\rm B} T_{\rm syst} \sqrt{\Delta_f/\tau}$, where for SHUKET $T_{\rm syst} = 554\;\rm K$, $\Delta_f$ is the filter width (1 Hz here) and $\tau$ is the measurement time (in our case 8000 seconds). This formula yields a detectable power of $8.6\times 10^{-23}\;\rm W$ in 1 Hz bins, to be compared to the measured $2\times 10^{-22}$ W. The difference is likely due to gain fluctuation, a 2.5\% fluctuation is enough to account for the difference. The narrow-band 22.2-hour run leads to a similar values of power density, with reduced dispersion. In that case, the RMS of the distribution is $2.3\times 10^{-23}\;\rm W/Hz$ in a 50~kHz band centered on 6.331525~GHz.

\begin{figure}[h]
\includegraphics[width=.8\columnwidth]{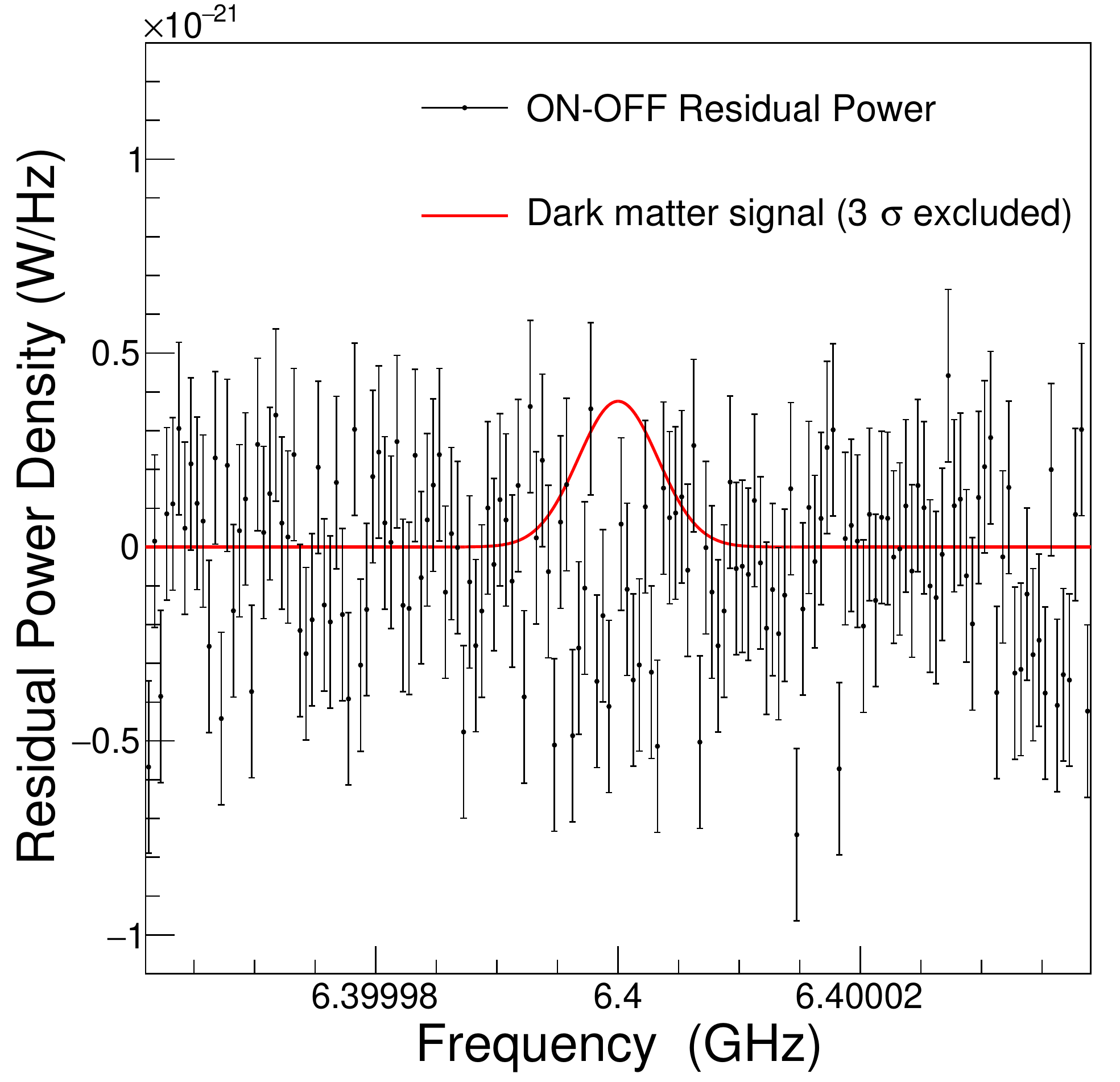}
\caption{Detail of the distribution of residuals around 6.4~GHz, together with a $3\; \sigma$-excluded dark matter signal.\label{fig:3}}
\end{figure}

A DM signal is searched for in the residual distribution. It is modeled as an excess power in the $ON$ distribution, with a gaussian spectral shape. The power carried out by the signal (Eq.~\ref{eq:power}) corresponds to the integral over frequencies. The width of the signal is fixed according to $\delta f / f = 5\times 10^{-7}$. Error bars are assigned to the data points along the following procedure: it is the RMS of the residual distribution measured on a sliding window of 1000 points. As an illustration, Fig.~\ref{fig:3} shows a zoom of the distribution of power residuals around 6.4~GHz, together with a  mock DM signal. In this specific example, the signal is excluded at the 3 $\sigma$ level.

\begin{figure}[h]
\includegraphics[width=.8\columnwidth]{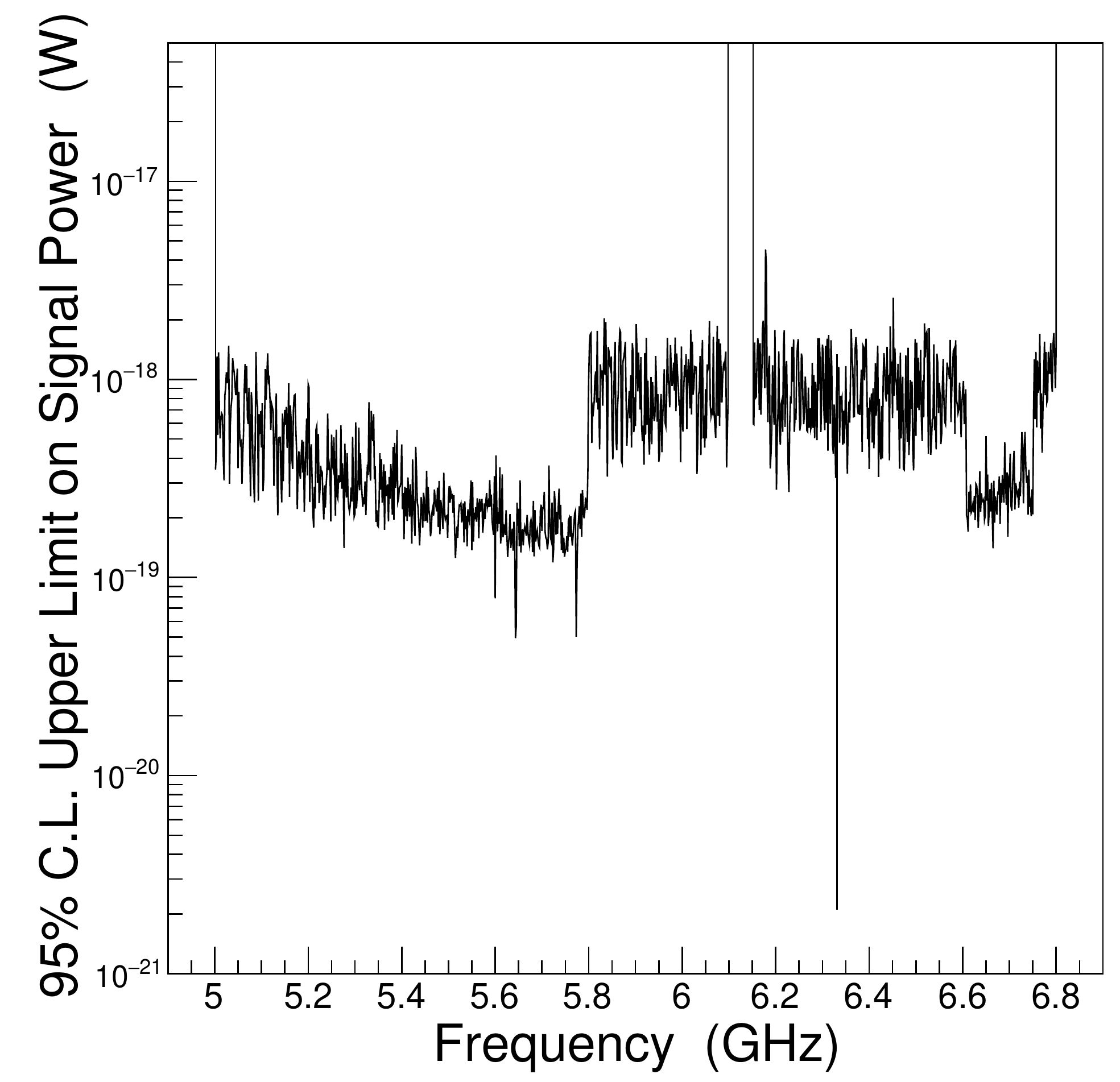}
\caption{95\% confidence level upper limits on the dark matter signal power. The vertical line at 6.331 GHz corresponds to the narrow-band search.\label{fig:4}}
\end{figure}

At each frequency, the presence of a residual power in the form of a DM signal is searched. No signal has been found. A $\chi^2$ analysis is performed to determine the confidence levels for the upper limit on the signal. The results are presented in Fig.~\ref{fig:4} at the 95\% confidence level (C.L.). The average exclusion on a DM signal power is of the order of $10^{-18}$ W. The drift of the sensitivity below 5.8 GHz and around 6.6 GHz is explained by a small shift downwards in the $ON$ power distribution in the corresponding frequency range. The improved limit at 6.331525~GHz yields an upper limit of $2.1\times 10^{-21}$ W.

\begin{figure}[h]
\includegraphics[width=.8\columnwidth]{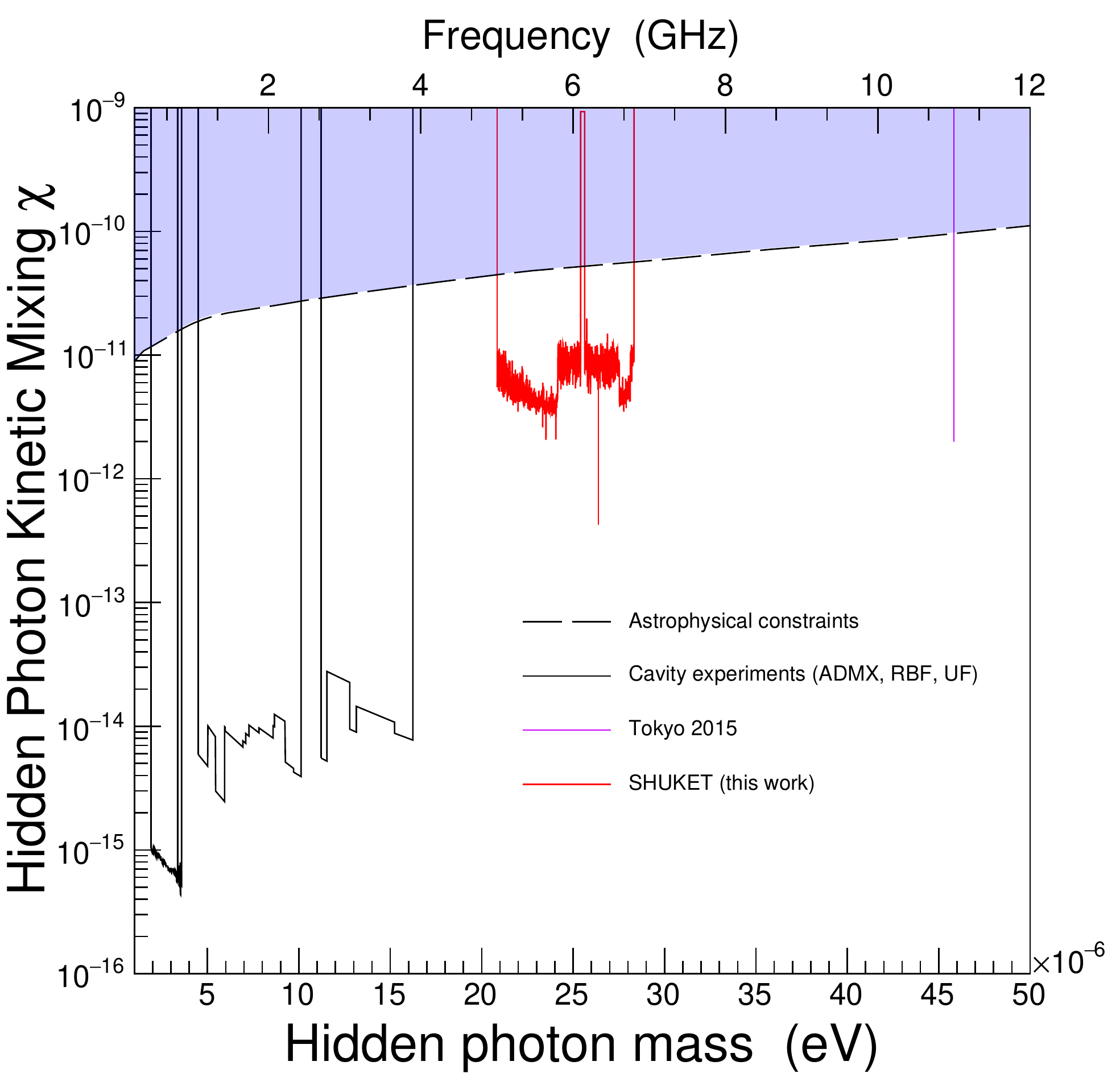}
\caption{95\% confidence level upper limits on the hidden photon kinetic mixing term (red curve).\label{fig:5}}
\end{figure}

{\it Interpretation --}
Equation~\ref{eq:power} is used to interpret the upper limits of Fig.~\ref{fig:4} in terms of the mixing parameter $\chi$. Assuming a typical value for the local DM density $\rho=0.3\;\rm GeV/cm^3$, and that hidden photons account for all of the DM, the upper limit on $\chi$ is of the order of $5\times 10^{-12}$ for HP masses between 20.8~$\mu$eV and  28.3~$\mu$eV. The results are presented in Fig.~\ref{fig:5}, together with other constraints. The blue region is excluded from the analysis of the cosmic microwave background (CMB) and the absence of anomalous distortions, see~\cite{2012JCAP...06..013A}. The exclusion regions on the left come from axion-search cavity experiments RBF~\cite{1987PhRvL..59..839D,Wuensch.1989sa}, UF~\cite{Hagmann.1990tj} and ADMX~\cite{2002ApJ...571L..27A,2010PhRvL.104d1301A}. The results from the axion cavity experiments are interpreted here in terms of HP following the method of~\cite{2012JCAP...06..013A}. The same hypothesis as above regarding the velocity distribution of the HP DM are applied. The purple line on the right is a narrow-band search for HP performed with a television antenna and a plane mirror~\cite{Suzuki.2015vka}.  Other similar searches have been performed in optical range using optical mirrors~\cite{2015JCAP...09..042S, 2017arXiv171102961E} sensitive to $\sim 10^{-10}-10^{-11}$ mixing parameter values, and at higher frequency with a plane mirror/dish system~\cite{2018JCAP...11..031K} sensitive to $\sim 10^{-8}$. Both these experiments yield constraints at  masses higher than $5\times 10^{-5}$ eV and lie outside the range of Fig.~\ref{fig:5}.

The SHUKET results are shown in red, both the wide-band and the narrow-band searches are included in Fig. 5. In string theory models with compactifications of heterotic orbifold, the mixing parameter is expected to be of the order of $\chi~\sim10^{-3}$~\cite{1997NuPhB.492..104D, 2012JHEP...01..021G}. In that case our results severely constrain the DM density in the mass range. In that case, and if the HP mass is in the range of sensitivity of SHUKET, the proportion of DM of that type would be less than $10^{-16}$. Should the value of the mixing parameter be $\chi=10^{-11}$ as suggested in~\cite{2009JHEP...11..027G} in the case of large volume string compactification, our results would imply this type of DM contribute to less than 44\% to the local DM density.

{\it Summary --}
We built an experiment dedicated to the search for hidden photons in the mass range 20.8~$\mu$eV to  28.3~$\mu$eV. Given the sensitivity to an excess power of the order of $10^{-22}\;\rm W/Hz$ in a frequency band spanning from 5 GHz to 6.8 GHz and the absence of anomalous signals, we have been able to exclude values of a power excess above $10^{-18}$ W. This excludes values of the kinetic mixing parameter above $5\times 10^{-12}$ at the 95\% C.L. for hidden photon masses between 20.8~$\mu$eV and  28.3~$\mu$eV. This mass range is particularly well motivated for axion dark matter, so a possible upgrade of SHUKET would be to magnetize the dish in order to be sensitive to axions as well.

\label{sec:phys}

{\it Acknowledgements --}
Part of this work was supported by the French national program PNHE and the ANR project CosmoTeV.
We acknowledge support from Rohde \& Schwarz. We would like to thank Jean-François Glicenstein, Fran\c{c}ois Brun, and Philippe Brax for careful reading and help in improving this Letter, Ahmimed Ouraou for his help with statistical analyses and Christophe Magneville for instructive discussions on radio astronomy. We are very grateful to Christian Herviou for providing us with the calibrated antenna and the shielding.

\bibliography{BCF}

\end{document}